\documentclass{PoS}
\usepackage{amssymb}
\usepackage{amsmath}
\usepackage{graphicx}

\newcommand{\ii}{\mathrm{i}}

\newcommand{\lnc}{\Lambda_{\mathrm{NC}}}

\PoS{PoS(HEP2005)322}

\title{Collider Tests of the Non-Commutative Standard Model}

\ShortTitle{Collider Tests of the Non-Commutative Standard Model}

\author{Ana Alboteanu\\
        University of W\"urzburg, Germany\\
        E-mail: \email{alboteanu@physik.uni-wuerzburg.de}}
\author{Thorsten Ohl\\
        University of W\"urzburg, Germany\\
        E-mail: \email{ohl@physik.uni-wuerzburg.de}}

\author{\speaker{Reinhold R\"uckl}\\
        University of W\"urzburg, Germany\\
        E-mail: \email{rueckl@physik.uni-wuerzburg.de}}

\abstract{We discuss the non-commutative extension of the 
  standard model constructed with the Seiberg-Witten maps for 
  $\mathrm{SU}(3)_C\otimes \mathrm{SU}(2)_L\otimes \mathrm{U}(1)_Y$.
  Using the first-order approximation in the non-commutative parameters 
  $\theta^{\mu\nu}$, we estimate the sensitivity of $Z\gamma$-production 
  at the Tevatron and LHC from Monte Carlo simulation.}

\FullConference{International Europhysics Conference on High Energy Physics\\
                 July 21st - 27th 2005\\
                 Lisboa, Portugal}

\begin{document}

\section{Gauge Theory on Non-Commutative Space-Time}
Ever since it was shown that quantum field theory on a
non-commutative (NC) space-time arises naturally in the low energy
limit of string theory \cite{Seiberg:1999vs}, there has been a lot of
interest in building realistic models, working out the phenomenology
and testing the predictions by experiment.

NC space-time is characterized by a nonvanishing commutator among coordinates,
\begin{equation}
\label{eq:commutator}
  [\hat{x}^\mu,\hat{x}^\nu] = \textrm{i}\theta^{\mu\nu}=
  \textrm{i}\frac{C^{\mu\nu}}{\lnc^2}\,,
\end{equation}
with a typical energy or inverse length scale $\lnc$. The current bounds
on $\lnc$ are strongly model dependent, ranging from
141\,GeV in collider experiments up to $10^{14}$\,GeV from
rotation invariance tests in atomic physics and astrophysics~\cite{bounds}.
In the following, we will assume the matrix $C^{\mu\nu}$ to be
constant and
realize the commutator~(\ref{eq:commutator}) by replacing the product
of functions of the NC coordinates $f(\hat x)\cdot g(\hat x)$ by the
Moyal-Weyl $\star$-product of functions of the ordinary
space-time coordinates:
\begin{equation}
  (f\star g)(x) = f(x) \exp(\frac{\ii}{2}\overleftarrow{\partial_\mu}
        \theta^{\mu\nu} \overrightarrow{\partial_\nu}) g(x)\,.
\end{equation}
In the case of gauge theories, this approach is only consistent for
$\mathrm{U}(N)$ gauge groups and only a single eigenvalue is allowed
for the charge operator of each $\mathrm{U}(1)$ factor.  For a NC
extension of the standard model (SM), the construction must therefore be
amended.  A minimal solution is provided by the Seiberg-Witten maps
(SWM) \cite{Seiberg:1999vs}, that express the NC gauge and matter
fields as non-linear functions of ordinary gauge and matter fields, such
that the gauge transformations of the non-commutative fields are
realized by the gauge transformations of the ordinary fields: $\hat
A'(A)=\hat A(A')$ and $\hat \psi'(\psi,A)=\hat \psi(\psi',A')$.  The
SWM can be determined from these conditions order by order
in~$\theta^{\mu\nu}$ for arbitrary gauge groups and representations.
In~\cite{Calmet:2001na} the NC generalisation of the $\mathrm{SU}(3)_C\times
\mathrm{SU}(2)_L\times \mathrm{U}(1)_Y$ standard model (NCSM) has been
constructed in first order in $\theta^{\mu\nu}$.

\section{Cross Section for $f\bar f\to Z\gamma$}
Starting from the Lagrangian in~\cite{Calmet:2001na}, we determined the
relevant Feynman rules in the NCSM (cf.~\cite{Ohl/Reuter:2004:NCPC}).
Interaction vertices that are already present in the SM acquire
momentum dependent corrections and new vertices like contact terms and
couplings among neutral gauge boson appear.
All these effects are present in $f\bar f\to Z\gamma$, the amplitude
of which is given by $A=A^{\text{SM}}+A^{\text{NC}}$, where, at $O(\theta^{\mu\nu})$,
\begin{equation*}
  A^{\text{NC}} =
    \parbox{45\unitlength}{\includegraphics{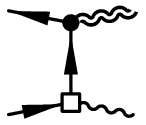}} +
    \parbox{45\unitlength}{\includegraphics{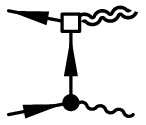}} +
    \parbox{45\unitlength}{\includegraphics{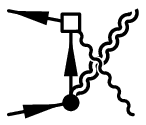}} +
    \parbox{45\unitlength}{\includegraphics{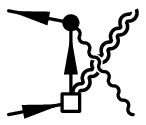}} +
    \parbox{45\unitlength}{\includegraphics{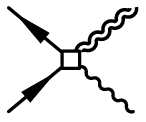}} +
    \parbox{45\unitlength}{\includegraphics{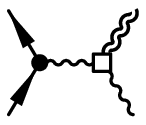}} +
    \parbox{45\unitlength}{\includegraphics{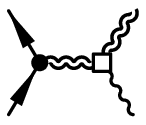}} \,.
\end{equation*}
The new and modified vertices have been marked by an open box in the Feynman
diagrams above.
The coupling strength of the contact term is fixed by gauge invariance,
while the $s$-channel diagrams depend on two new coupling
constants~$K_{Z\gamma\gamma}$ and~$K_{ZZ\gamma}$ that are only
constrained by the matching to the SM in the limit
$\theta^{\mu\nu}\to0$ and can vary independently in a finite
range~\cite{Calmet:2001na}.

\section{%
  Sensitivity of the channel $q\bar q\to e^+e^-\gamma$
  at the Tevatron and LHC}
We have performed a phenomenological analysis of the NC effects at
hadron colliders in the process $q\bar q\to e^+e^-\gamma$, which
includes the resonant $Z\gamma$ contribution.  The complete $q\bar q\to
e^+e^-\gamma$ cross section was calculated numerically using an
unpublished extension of the libraries of
O'Mega~\cite{Ohl/etal:Omega}. The Monte Carlo simulations were
performed with WHiZard \cite{Kilian:WHIZARD}.

\begin{figure}
  \centering
  \includegraphics[width=0.7\textwidth]{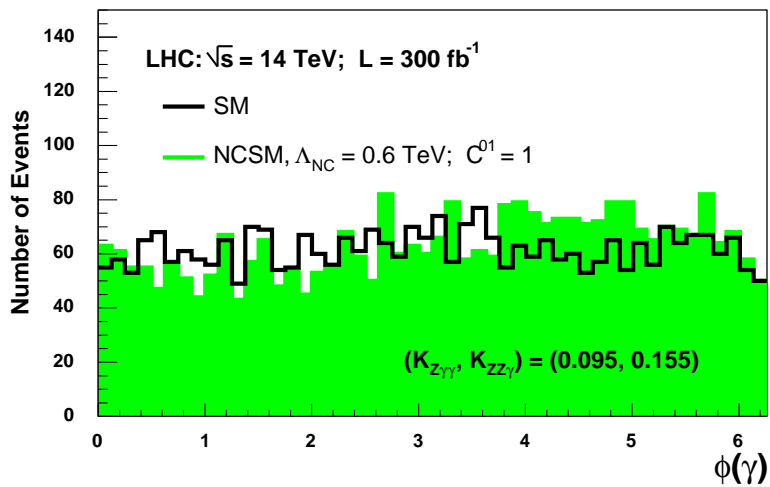}
  \caption{\label{fig:azimuthal}%
    Azimuthal distribution of the photon in $pp\to e^+e^-\gamma$
    at the LHC, applying the cuts $\cos\theta_{e^+e^-} > 0$, $\cos\theta_\gamma > 0$
    and $0<\cos\theta^*_\gamma<0.9$.}
\end{figure}

Cross sections in the NCSM show a characteristic dependence on the
azimuthal angle~$\phi$, which can be used to search for NC signals
and to discriminate against other new physics effects.  However, this
effect is linear in the partonic $\cos\theta^*$ and cancels at the LHC
due to the symmetric initial state $pp$, even if we require $\cos\theta^*>0$.
Nevertheless, since the average momentum fraction of quarks in the
proton is much higher than the average momentum fraction of
antiquarks, one can select either $q\bar q\to Z\gamma$ or $\bar q q\to Z\gamma$
events at the LHC, by requiring a minimal boost of the partonic center of
mass system in the direction of the quark.  A simple and efficient cut is to require both the
$\gamma$ and the $e^+e^-$-pair to be produced in the same hemisphere.
The resulting distribution at the LHC is depicted in fig.~\ref{fig:azimuthal}.
At the Tevatron, the corresponding events with a high
partonic~$\sqrt{\hat s}$ are mostly $q\bar q\to Z\gamma$ without additional cuts.
However, the NC effect is much weaker~\cite{Alboteanu/etal:NCLHC}.

\begin{figure}
  \begin{center}
    \includegraphics[width=0.24\textwidth]{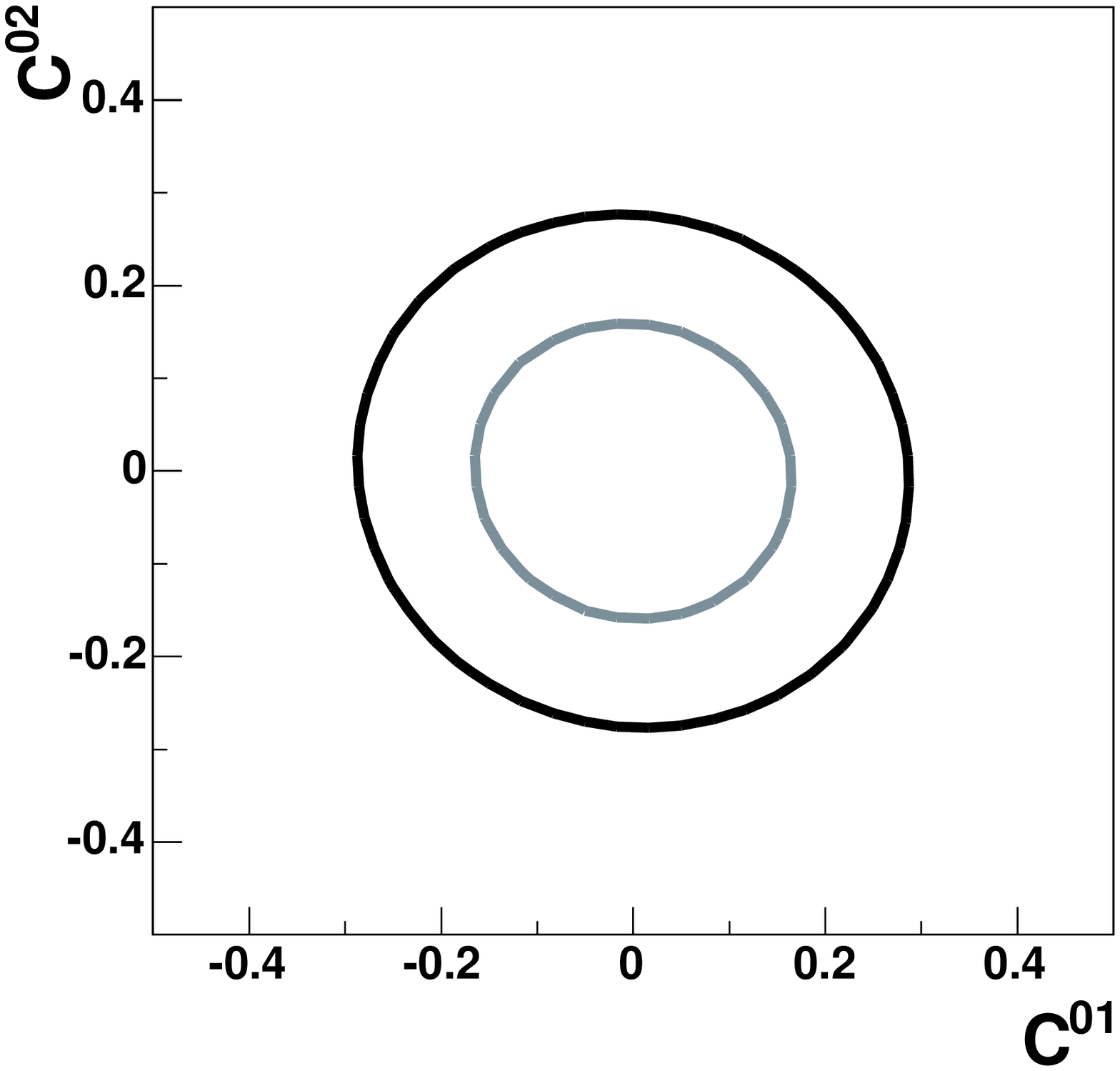} 
    \includegraphics[width=0.24\textwidth]{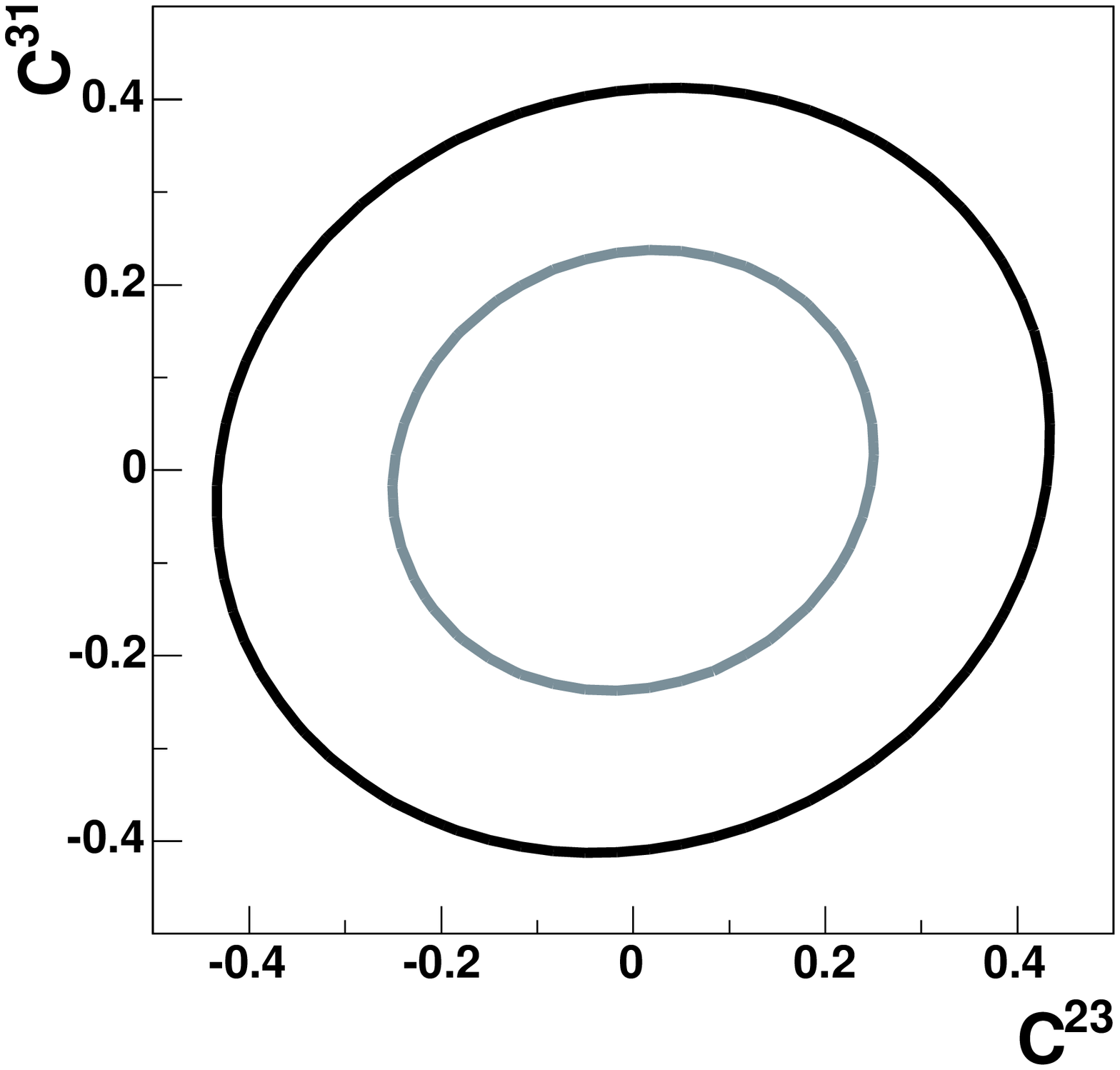} 
    \includegraphics[width=0.24\textwidth]{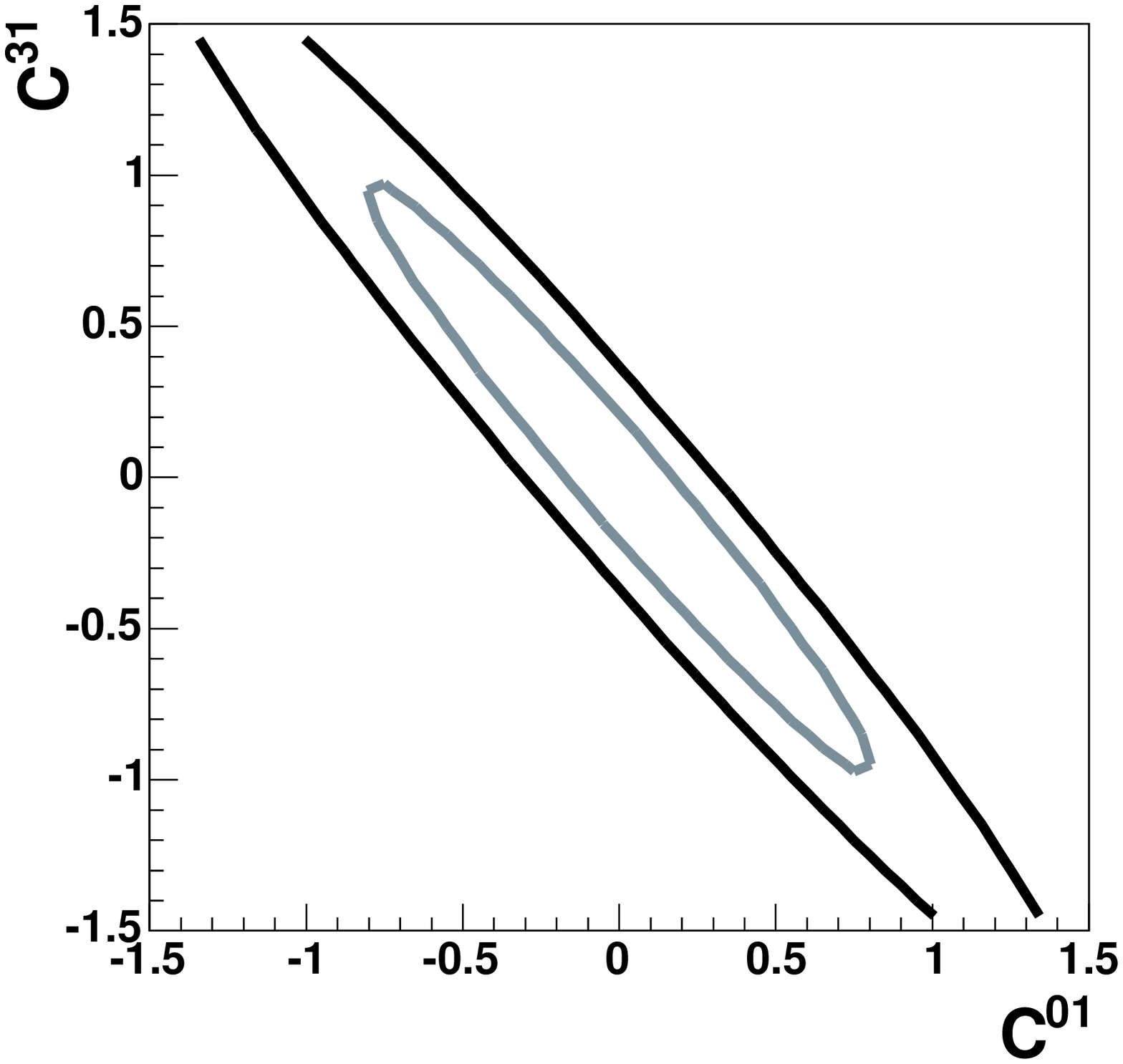}
    \includegraphics[width=0.24\textwidth]{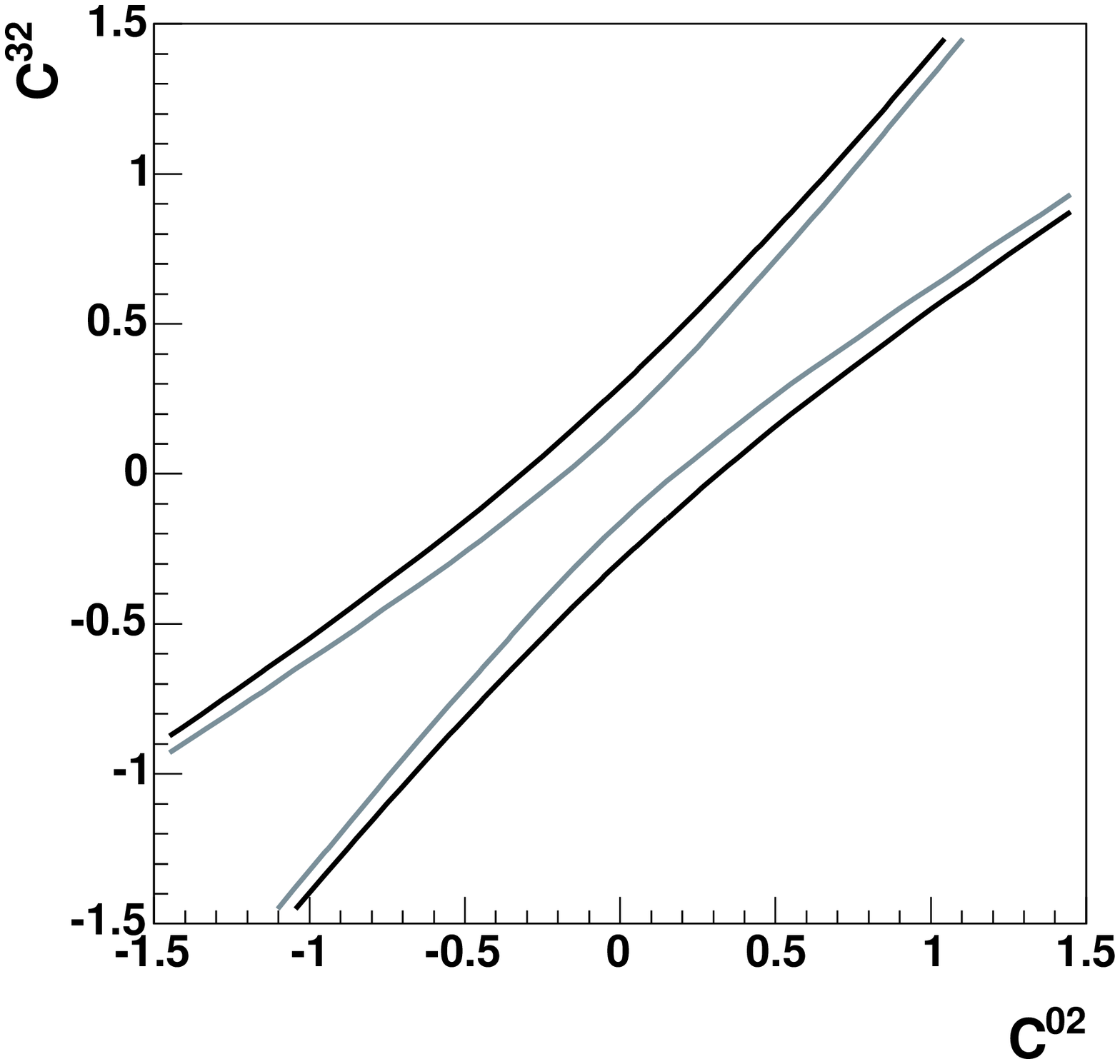}
  \end{center}
  \caption{\label{fig:contours}%
    Sensitivity on $C^{\mu\nu}$ at the LHC for an integrated
    luminosity of $100\,\textrm{fb}^{-1}$ assuming
    $\lnc=500\,\text{GeV}$ and applying the kinematical cuts given in
    fig.~\protect\ref{fig:azimuthal}.}
\end{figure}

Using the azimuthal dependence of the cross section in first order in
$\theta^{\mu\nu}$ plotted in fig.~\ref{fig:azimuthal}, we have
estimated the sensitivity on the
scale $\lnc$ from fitting the quadratic dependence of~$\chi^2$ on
$\theta^{\mu\nu}$.  The resulting error ellipses for~$C^{\mu\nu}$ at
fixed $\lnc=500\,\text{GeV}$ are shown in fig.~\ref{fig:contours}.
The correlations between $(C^{01}, C^{13})$ and
$(C^{02},C^{23})$, respectively, are expected from kinematics.
In the center of mass system, the partonic cross sections show a much
stronger dependence on the
time-like components $C^{0i}$ than on the space-like components
$C^{ij}$.  However, in the
laboratory frame, the time-like and
space-like components are mixed by Lorentz boosts, e.\,g.
$C^{01} \to \gamma (C^{01} - \beta C^{13})$, leading to the strong
correlations observed in fig.~\ref{fig:contours}.
Analytically, the corresponding contours would be straight
diagonal lines.  The depicted stretched ellipse and hyperbola are the
result of fluctuations of the vanishing eigenvalues of the quadratic
form for~$\chi^2$, fitted from the simulated event samples.

From the error ellipses in fig.~\ref{fig:contours},
we derive the following sensitivity limits on the NC
scale at the LHC for an integrated luminosity
of~$100\,\textrm{fb}^{-1}$:
\begin{equation}
  \lnc \geq 1.2 \ \text{TeV} \;\text{for}\; C^{0i}=1, \quad
  \lnc \geq 1.0 \ \text{TeV} \;\text{for}\; C^{3i}=1, \quad i = 1,2\,.
\end{equation}
Such a measurement would improve the current collider bounds by an
order of magnitude.  A similar analysis for the Tevatron shows that
the sensitivity on the NC scale remains below $100\,\textrm{GeV}$
there~\cite{Alboteanu/etal:NCLHC}.

Our analysis is based on the scattering amplitude
$A=A^{\text{SM}}+A^{\text{NC}}$ to first order in the
noncommutativity~$\theta^{\mu\nu}$.  Therefore the NC effects arise
only from the SM/NC interference term which scales with $\hat
s/{\lnc}^2$.  It turns out that values of $\lnc$ that are observable
in events with typical values of~$\sqrt{\hat s}$ can correspond to an
unphysical cross section at the highest~$\sqrt{\hat s}$ available.  We
have eliminated this problem by assuming that higher orders in
$\theta^{\mu\nu}$ will cancel the large negative interference
contributions.  Preliminary results for the calculation in second
order in $\theta^{\mu\nu}$ support this
prejudice~\cite{Alboteanu/etal:NCLHC}.  We also find that
the second order contributions to the unpolarized cross section do not
cancel for symmetric final states such as $\gamma\gamma$ as they did
in first order~\cite{Ohl/Reuter:2004:NCPC}.  This could provide
signatures for the NCSM also in $pp\to\gamma\gamma$, a process which
will be studied in great detail for Higgs searches at the LHC.

\end{document}